\documentclass[fleqn,twoside]{article}
\usepackage{espcrc2}

\newcommand{\be}{\begin{equation}}
\newcommand{\ee}{\end{equation}}

\newcommand{\ba}{\begin{eqnarray}}
\newcommand{\ea}{\end{eqnarray}}

\newcommand{\Tr}{{\rm Tr}}

\title{Infrared finitness and analyticity properties of the\\
       loop--loop scattering amplitudes in gauge theories}

\author{E. Meggiolaro\address{Dipartimento di Fisica,
        Universit\`a di Pisa, Largo Pontecorvo 3, I--56127 Pisa, Italy\\
	E-mail: enrico.meggiolaro@df.unipi.it
}}
\begin{document}

\begin{abstract}

We shall discuss about the infrared finitness and some analyticity properties
of the loop--loop scattering amplitudes in gauge theories, when going from
Minkowskian to Euclidean theory, and we shall see how they can be related to
the still unsolved problem of the $s$--dependence of the hadron--hadron total
cross--sections.

\end{abstract}

\maketitle

\noindent
Differently from the parton--parton scattering amplitudes, which are known to
be affected by infrared (IR) divergences, the elastic scattering amplitude of
two colourless states in gauge theories, e.g., two $q \bar{q}$ meson states,
is expected to be an IR--finite physical quantity.
It was shown in Refs. \cite{Nachtmann97,Dosch,Berger} that the high--energy
meson--meson elastic scattering amplitude can be approximately reconstructed
by first evaluating, in the eikonal approximation, the elastic scattering
amplitude of two $q \bar{q}$ pairs (usually called ``{\it dipoles}''), of
given transverse sizes $\vec{R}_{1\perp}$ and $\vec{R}_{2\perp}$ respectively,
and then averaging this amplitude over all possible values of
$\vec{R}_{1\perp}$ and $\vec{R}_{2\perp}$ with two proper squared
wave functions $|\psi_1 (\vec{R}_{1\perp})|^2$ and
$|\psi_2 (\vec{R}_{2\perp})|^2$, describing the two interacting mesons.
The high--energy elastic scattering amplitude of two {\it dipoles}
is governed by the correlation function of two Wilson loops
${\cal W}_1$ and ${\cal W}_2$, which follow the classical straight lines for
quark (antiquark) trajectories:
\ba
\lefteqn{
{\cal T}_{(ll)} (s,t;~\vec{R}_{1\perp},\vec{R}_{2\perp}) \equiv }
\nonumber \\
& & -i~2s \displaystyle\int d^2 \vec{z}_\perp
e^{i \vec{q}_\perp \cdot \vec{z}_\perp}
\left[ {\langle {\cal W}_1 {\cal W}_2 \rangle \over
\langle {\cal W}_1 \rangle \langle {\cal W}_2 \rangle} -1 \right] ,
\label{scatt-loop}
\ea
where $s$ and $t = -\vec{q}_\perp^2$ ($\vec{q}_\perp$ being the tranferred
momentum) are the usual Mandelstam variables.
More explicitly the Wilson loops ${\cal W}_1$ and ${\cal W}_2$ are so defined:
\ba
\lefteqn{
{\cal W}^{(T)}_1 \equiv
{1 \over N_c} \Tr \left\{ {\cal P} \exp
\left[ -ig \displaystyle\oint_{{\cal C}_1} A_\mu(x) dx^\mu \right] \right\} , }
\nonumber \\
\lefteqn{
{\cal W}^{(T)}_2 \equiv
{1 \over N_c} \Tr \left\{ {\cal P} \exp
\left[ -ig \displaystyle\oint_{{\cal C}_2} A_\mu(x) dx^\mu \right] \right\} , }
\label{QCDloops}
\ea
where ${\cal P}$ denotes the ``{\it path ordering}'' along the given path
${\cal C}$; ${\cal C}_1$ and ${\cal C}_2$ are two rectangular paths which
follow the classical straight lines for quark [$X_{(+)}(\tau)$, forward in
proper time $\tau$] and antiquark [$X_{(-)}(\tau)$, backward in $\tau$]
trajectories, i.e.,
\ba
{\cal C}_1 &\to&
X_{(\pm 1)}^\mu(\tau) = z^\mu + {p_1^\mu \over m} \tau
\pm {R_1^\mu \over 2} , \nonumber \\
{\cal C}_2 &\to&
X_{(\pm 2)}^\mu(\tau) = {p_2^\mu \over m} \tau \pm {R_2^\mu \over 2} ,
\label{traj}
\ea
and are closed by straight--line paths at proper times $\tau = \pm T$, where
$T$ plays the role of an IR cutoff, which must be removed at the end
($T \to \infty$).
Here $p_1$ and $p_2$ are the four--momenta of the two quarks and of the two
antiquarks with mass $m$, moving with speed $\beta$ and $-\beta$ along, for
example, the $x^1$--direction:
\ba
p_1 &=& m (\cosh {\chi \over 2},\sinh {\chi \over 2},0,0) ,
\nonumber \\
p_2 &=& m (\cosh {\chi \over 2},-\sinh {\chi \over 2},0,0) ,
\label{p1p2}
\ea
where $\chi = 2~{\rm arctanh} \beta$ is the hyperbolic angle between the two
trajectories $(+1)$ and $(+2)$.
Moreover, $R_1 = (0,0,\vec{R}_{1\perp})$, $R_2 = (0,0,\vec{R}_{2\perp})$
and $z = (0,0,\vec{z}_\perp)$, where $\vec{z}_\perp = (z^2,z^3)$ is the
impact--parameter distance between the two loops in the transverse plane.

It is convenient to consider also
the correlation function of two Euclidean Wilson loops
$\tilde{\cal W}_1$ and $\tilde{\cal W}_2$ running along two rectangular paths
$\tilde{\cal C}_1$ and $\tilde{\cal C}_2$ which follow the following
straight--line trajectories:
\ba
\tilde{\cal C}_1 &\to&
X^{(\pm 1)}_{E\mu}(\tau) = z_{E\mu} + {p_{1E\mu} \over m}
\tau \pm {R_{1E\mu} \over 2} , \nonumber \\
\tilde{\cal C}_2 &\to&
X^{(\pm 2)}_{E\mu}(\tau) = {p_{2E\mu} \over m} \tau
\pm {R_{2E\mu} \over 2} ,
\label{trajE}
\ea
and are closed by straight--line paths at proper times $\tau = \pm T$. Here
$R_{1E} = (0,\vec{R}_{1\perp},0)$, $R_{2E} = (0,\vec{R}_{2\perp},0)$ and
$z_E = (0,\vec{z}_\perp,0)$. Moreover, in the Euclidean theory we {\it choose}
the four--vectors $p_{1E}$ and $p_{2E}$ to be:
\ba
p_{1E} &=& m (\sin{\theta \over 2}, 0, 0, \cos{\theta \over 2} ) , \nonumber \\
p_{2E} &=& m (-\sin{\theta \over 2}, 0, 0, \cos{\theta \over 2} ) ,
\label{p1p2E}
\ea
where $\theta \in [0,\pi]$ is the angle formed by the two trajectories
$(+1)$ and $(+2)$
in Euclidean four--space.\\
Let us introduce the following notations for the normalized correlators
$\langle {\cal W}_1 {\cal W}_2 \rangle / \langle {\cal W}_1 \rangle
\langle {\cal W}_2 \rangle$ in the Minkowskian and in the Euclidean theory,
in the presence of a {\it finite} IR cutoff $T$:
\ba
\label{GM-GE}
\lefteqn{
{\cal G}_M(\chi;~T;~\vec{z}_\perp,\vec{R}_{1\perp},\vec{R}_{2\perp}) \equiv
{ \langle {\cal W}^{(T)}_1 {\cal W}^{(T)}_2 \rangle \over
\langle {\cal W}^{(T)}_1 \rangle
\langle {\cal W}^{(T)}_2 \rangle } ,} \\ 
\lefteqn{
{\cal G}_E(\theta;~T;~\vec{z}_\perp,\vec{R}_{1\perp},\vec{R}_{2\perp}) \equiv
{ \langle \tilde{\cal W}^{(T)}_1 \tilde{\cal W}^{(T)}_2 \rangle_E \over
\langle \tilde{\cal W}^{(T)}_1 \rangle_E
\langle \tilde{\cal W}^{(T)}_2 \rangle_E } . } \nonumber
\ea
As already stated in Ref. \cite{Meggiolaro02}, the two quantities in Eq.
(\ref{GM-GE}) are expected to be connected by the same analytic continuation
in the angular variables and in the IR cutoff which was already derived in the
case of Wilson lines \cite{Meggiolaro02,Meggiolaro97,Meggiolaro98}, i.e.:
\ba
\lefteqn{
{\cal G}_M(\chi;~T;~\vec{z}_\perp,\vec{R}_{1\perp},\vec{R}_{2\perp}) = }
\nonumber \\
& & {\cal G}_E(\theta \to -i\chi;~T \to iT;
~\vec{z}_\perp,\vec{R}_{1\perp},\vec{R}_{2\perp}) .
\label{analytic}
\ea
Indeed it can be proved \cite{Meggiolaro04}, simply by adapting step by step
the proof derived in Ref.  \cite{Meggiolaro02} from the case of Wilson lines to
the case of Wilson loops, that the analytic continuation (\ref{analytic}) is an
{\it exact} result, i.e., not restricted to some order in perturbation theory
or to some other approximation, and is valid both for the Abelian and the
non--Abelian case.

As we have said above, the loop--loop correlation functions (\ref{GM-GE}),
both in the Minkowskian and in the Euclidean theory, are expected to be
IR--{\it finite} quantities, i.e., to have finite limits when $T \to \infty$,
differently from what happens in the case of Wilson lines.
One can then define the following loop--loop correlation function in the
Minkowskian theory with the IR cutoff removed,
\ba
\lefteqn{
{\cal C}_M(\chi;~\vec{z}_\perp,\vec{R}_{1\perp},\vec{R}_{2\perp}) \equiv }
\nonumber \\
& & \displaystyle\lim_{T \to \infty} \left[
{\cal G}_M(\chi;~T;~\vec{z}_\perp,\vec{R}_{1\perp},\vec{R}_{2\perp})
- 1 \right] ,
\label{C12}
\ea
and the corresponding quantity in the Euclidean theory,
${\cal C}_E(\theta;~\vec{z}_\perp,\vec{R}_{1\perp},\vec{R}_{2\perp})$.

As a pedagogic example to illustrate these considerations, we shall consider
the simple case of QED, in the so--called {\it quenched} approximation, where
vacuum polarization effects, arising from the presence of loops of dynamical
fermions, are neglected.
In this approximation, the calculation of the normalized correlators
(\ref{GM-GE}) can be performed exactly (i.e., without further approximations)
both in Minkowskian and in Euclidean theory and one finds that
\cite{Meggiolaro04} i) the two quantities ${\cal G}_M$ and ${\cal G}_E$ are
indeed connected by the analytic continuation (\ref{analytic}), and ii) the
two quantities are finite in the limit when the IR cutoff $T$ goes to
infinity:
\ba
\lefteqn{
{\cal C}_M(\chi;~\vec{z}_\perp,\vec{R}_{1\perp},\vec{R}_{2\perp}) = }
\nonumber \\
& & \exp \left[ -i 4e^2 \coth \chi~
t(\vec{z}_\perp,\vec{R}_{1\perp},\vec{R}_{2\perp}) \right] - 1 ,
\label{QED-M} \\
\lefteqn{
{\cal C}_E(\theta;~\vec{z}_\perp,\vec{R}_{1\perp},\vec{R}_{2\perp}) = }
\nonumber \\
& & \exp \left[ - 4e^2 \cot \theta~
t(\vec{z}_\perp,\vec{R}_{1\perp},\vec{R}_{2\perp}) \right] - 1 ,
\label{QED-E}
\ea
where
\ba
\lefteqn{
t(\vec{z}_\perp,\vec{R}_{1\perp},\vec{R}_{2\perp}) \equiv }
\nonumber \\
\lefteqn{
{1 \over 8\pi} \ln \left(
{ |\vec{z}_\perp+{\vec{R}_{1\perp} \over 2}+{\vec{R}_{2\perp} \over 2}|
  |\vec{z}_\perp-{\vec{R}_{1\perp} \over 2}-{\vec{R}_{2\perp} \over 2}| \over
  |\vec{z}_\perp+{\vec{R}_{1\perp} \over 2}-{\vec{R}_{2\perp} \over 2}|
  |\vec{z}_\perp-{\vec{R}_{1\perp} \over 2}+{\vec{R}_{2\perp} \over 2}| }
\right) }
\label{t-function}
\ea
As shown in Ref. \cite{Meggiolaro04}, the results (\ref{QED-M}) and
(\ref{QED-E}) can be used to derive the corresponding results in the case of a
non--Abelian gauge theory with $N_c$ colours, up to the order ${\cal O}(g^4)$
in perturbation theory (see also Refs. \cite{LLCM,BB}):
\ba
\lefteqn{
{\cal C}_M(\chi;~\vec{z}_\perp,\vec{R}_{1\perp},\vec{R}_{2\perp})|_{g^4} = }
\nonumber \\
\lefteqn{
- 2g^4 \left( {N_c^2 - 1 \over N_c^2} \right) \coth^2 \chi~
[t(\vec{z}_\perp,\vec{R}_{1\perp},\vec{R}_{2\perp})]^2 , }
\label{QCD-pertM} \\
\lefteqn{
{\cal C}_E(\theta;~\vec{z}_\perp,\vec{R}_{1\perp},\vec{R}_{2\perp})|_{g^4} = }
\nonumber \\
\lefteqn{
2g^4 \left( {N_c^2 - 1 \over N_c^2} \right) \cot^2 \theta~
[t(\vec{z}_\perp,\vec{R}_{1\perp},\vec{R}_{2\perp})]^2 . }
\label{QCD-pertE}
\ea

We stress the fact that both the Minkowskian quantities (\ref{QED-M}) and
(\ref{QCD-pertM}) and the Euclidean quantities (\ref{QED-E}) and
(\ref{QCD-pertE}) are IR finite when $T \to \infty$, differently from
the corresponding quantities constructed with Wilson lines,
which were evaluated in Ref. \cite{Meggiolaro97} (see also Ref.
\cite{Meggiolaro96}).

It is also important to notice that the two quantities (\ref{QED-M}) and
(\ref{QED-E}), as well as the two quantities (\ref{QCD-pertM}) and
(\ref{QCD-pertE}), obtained {\it after} the removal of the IR cutoff
($T \to \infty$), are still connected by the usual analytic continuation in
the angular variables only:
\ba
\lefteqn{
{\cal C}_M(\chi;~\vec{z}_\perp,\vec{R}_{1\perp},\vec{R}_{2\perp}) = }
\nonumber \\
& & {\cal C}_E(\theta \to -i\chi;
~\vec{z}_\perp,\vec{R}_{1\perp},\vec{R}_{2\perp}) .
\label{final}
\ea
This is a highly non--trivial result, whose general validity is discussed
in Ref. \cite{Meggiolaro04}.
(Indeed, the validity of the relation (\ref{final}) has
been also recently verified in Ref. \cite{BB} by an explicit calculation
up to the order ${\cal O}(g^6)$ in perturbation theory.)\\
As said in Ref. \cite{Meggiolaro04},
if ${\cal G}_M$ and ${\cal G}_E$ are analytic functions of $T$ in the
whole complex plane and if $T=\infty$ is an ``eliminable singular point''
[i.e., the finite limit (\ref{C12}) exists when letting the {\it complex}
variable $T \to \infty$], then, of course, the analytic continuation
(\ref{final}) immediately derives from Eq. (\ref{analytic}), when letting
$T \to +\infty$. (For example, if ${\cal G}_M$ and ${\cal G}_E$ are analytic
functions of $T$ and they are bounded at large $|T|$, then $T=\infty$
is an ``eliminable singular point'' for both of them.)
But the same result (\ref{final}) can also be derived under weaker
conditions. For example, let us assume that ${\cal G}_E$ is a bounded
analytic function of $T$ in the sector $0 \le \arg T \le {\pi \over 2}$,
with finite limits along the two straight lines on the border of the sector:
${\cal G}_E \to G_{E1}$, for $({\rm Re}T \to +\infty,~{\rm Im}T = 0)$, and
${\cal G}_E \to G_{E2}$, for $({\rm Re}T = 0,~{\rm Im}T \to +\infty)$.
And, similarly, let us assume that ${\cal G}_M$ is a bounded
analytic function of $T$ in the sector $-{\pi \over 2} \le \arg T \le 0$,
with finite limits along the two straight lines on the border of the sector:
${\cal G}_M \to G_{M1}$, for $({\rm Re}T \to +\infty,~{\rm Im}T = 0)$, and
${\cal G}_M \to G_{M2}$, for $({\rm Re}T = 0,~{\rm Im}T \to -\infty)$.
We can then apply the ``Phragm\'en--Lindel\"of theorem''
to state that $G_{E2} = G_{E1}$ and
$G_{M2} = G_{M1}$. Therefore, also in this case, the analytic continuation
(\ref{final}) immediately derives from Eq. (\ref{analytic}) when
$T \to \infty$.

The relation (\ref{final}) has been extensively used in the literature in order
to address, from a non--perturbative point of view, the still unsolved problem
of the asymptotic $s$--dependence of hadron--hadron elastic scattering
amplitudes and total cross sections \cite{LLCM,JP,Janik,instanton1,instanton2}.
(It has been also recently proved in Ref. \cite{BB}, by an explicit
perturbative calculation, that the loop--loop scattering amplitude approaches,
at sufficiently high energy, the BFKL--{\it pomeron} behaviour \cite{BFKL}.)

An independent non--perturbative approach would be surely welcome and could be
provided by a direct lattice calculation of the loop--loop Euclidean
correlation functions.
This would surely result in a considerable progress along this line
of research.

\end{document}